\documentclass[twocolumn,prb,preprintnumbers,amsmath,amssymb,floatfix,showpacs]{revtex4}

\usepackage{graphicx}
\usepackage{subfigure}
\usepackage{dcolumn}

\begin{document}

\title{Anisotropy and Magnetism in the LSDA+U Method}
\author{Erik R. Ylvisaker}
\affiliation{Department of Physics, University of California, Davis, California, 95616}

\author{Warren E. Pickett}
\affiliation{Department of Physics, University of California, Davis, California, 95616}

\author{Klaus Koepernik}
\affiliation{IFW Dresden, P O Box 270116, D-01171 Dresden, Germany}

\date{\today}
%
\begin{abstract}

Consequences of anisotropy (variation of orbital occupation) and 
magnetism, and their coupling, are analyzed for LSDA+U functionals, 
both the commonly used ones as well as less commonly applied functionals.  
After reviewing and extending some earlier observations for an
isotropic interaction, the anisotropies are examined more fully and
related to use with the local density (LDA) or local spin density
(LSDA) approximations.  
The total energies of all possible integer configurations of an
open $f$ shell are presented for three functionals, where some
differences are found to be dramatic.  
Differences in how the commonly used ``around mean field'' (AMF) and ``fully
localized limit'' (FLL) functionals 
perform are traced to such differences.
The LSDA+U interaction term, applied self-consistently, usually enhances
spin magnetic moments and orbital polarization, and the double-counting terms of both functionals 
provide an opposing, moderating tendency (``suppressing the magnetic 
moment''). The AMF double counting term gives magnetic states a significantly 
larger energy penalty than does the FLL counterpart.

\end{abstract}


\maketitle

\section{Introduction}

Density functional theory (DFT) and its associated local (spin) density 
approximation [L(S)DA] is used widely to describe the properties of a 
wide variety of materials, often with great success.
However there exists a class of materials which are poorly described, 
sometimes qualitatively, by 
LDA. These so-called strongly correlated materials
typically contain atoms with open $d$ or $f$ shells, in which the
corresponding orbitals are in some sense localized.  
The LSDA+U approach was introduced 
by Anisimov, Zaanen, and Andersen\cite{OriginalLDAU} to treat correlated 
materials as a modification of LDA (`on top of LDA') that adds an 
intra-atomic Hubbard U repulsion term in the energy functional.  Treated in
a self-consistent mean field (`Hartree-Fock') manner, in quite a large number
of cases the LDA+U result provides a greatly improved description of 
strongly correlated materials.  

At the most basic level, the LSDA+U correction tends to drive the correlated
orbital $m$ occupation numbers $n_{m\sigma}$ ($\sigma$ denotes spin projection)
to integer values 0 or 1.  This in turn produces, under appropriate conditions,
insulating states out of conducting LSDA states, and the Mott insulating
state of several systems is regarded as being well described by LSDA+U at
the band theory level.   Dudarev {\it et al.}\cite{Dudarev98} and Petukhov
{\it et al.}\cite{Petukhov03} provided some description of the
effect of the spin dependence of two different double counting terms 
within an isotropic approximation.
Beyond this important but simple effect, there is 
freedom in which of the spin-orbitals ($m\sigma$) will be occupied, which can
affect the result considerably and therefore makes
it important to understand the effects of anisotropy and spin polarization in
LSDA+U.  After the successes of providing realistic pictures of the Mott
insulating state in La$_2$CuO$_4$ and the transition metal 
monoxides,\cite{OriginalLDAU} the anisotropy contained in the LSDA+U method 
produced the correct orbitally ordered magnetic arrangement for KCuF$_3$ that provided
an understanding of its magnetic behavior.\cite{SashaL95}

The anisotropy of the interaction, and its connection to the level of spin
polarization, is a topic that is gaining interest and importance.  One example
is in the LSDA+U description of the zero temperature Mott transition under 
pressure in the classic Mott insulator MnO.  The first transition under pressure
is predicted to
be\cite{Kasinathan07} an insulator-insulator (not insulator-metal) transition, with a
$S=\frac{5}{2} \rightarrow S=\frac{1}{2}$ moment collapse and a volume collapse.
The insulator-to-insulator aspect is surprising, but more surprising is the form of
moment collapse: each orbital remains singly occupied beyond the transition, but
the spins of electrons in two of the orbitals have flipped direction.  This type 
of moment collapse is totally unanticipated (and hence disbelieved by some), but
it is robust against crystal structure (occurring in both rocksalt and NiAs
structures) and against reasonable variation of the interaction strength.  Detailed analysis indicates it is
a product of the anisotropy of the LSDA+U interaction and the symmetry lowering due to
antiferromagnetic order.

Another unanticipated result was obtained\cite{KWL2004} in LaNiO$_2$, which is a metal
experimentally. This compound is also a metal in LSDA+U over a very large range of 
interaction strength $U$, rather than reverting to a Mott insulating Ni$^{1+}$
system which would be isovalent with CaCuO$_2$.  
For values of $U$ in the range expected to be
appropriate for the Ni ion in this oxide, the magnetic system consists of an
atomic singlet consisting of antialigned $d_{x^2-y^2}$ and $d_{z^2}$ spins
on each Ni ion.  Again the anisotropy of the interaction evidently plays a
crucial role in the result, with its effect being coupled thoroughly with 
band mixing effects.

The addition of a Hubbard U interaction also introduces the need for 
``double counting'' correction
terms in the energy functional, to account for the fact that the Coulomb
energy is already included (albeit more approximately) in the LSDA functional.
All double counting schemes subtract an averaged energy for the occupation
of a selected reference state depending only on $\{N_{\sigma}\}$,
which largely cancels the isotropic
interaction of the $E_{I}$ term Eq. (\ref{eq:LDAU-General}).
Several forms for these double-counting terms have been
proposed,\cite{OriginalLDAU,Anisimov93,CS} but primarily two are
commonly used.  These LDA+U functionals are most often referred to as around mean
field (AMF) and the fully localized limit (FLL), which is also referred to as
the atomic limit (AL).  The distinctions between these forms have attracted
some discussion, but without consideration of the full anisotropy of the
interaction.

The need for double-counting corrections is not unique to the LDA+U method; 
any other method that adds correlation terms to the LSDA functional, such 
as the dynamical LDA+DMFT (dynamical mean field theory) approach, also will 
require double-counting corrections.  This is an unfortunate consequence of LDA's 
success; LDA works too well, even in correlated systems where it usually 
gets interatomic charge balance reasonably, to just throw it away.\footnote{There
are numerous examples for strongly correlated ({\it heavy fermion}) metals where
the Fermi surface calculated within LDA is predicted as well as for more conventional
metals.  This surely requires that the charge balance between the various atoms
is accurate.}  The common
approach has been to use LSDA for correlated materials and include a 
double-counting correction.  There are techniques being developed which do 
not build on a correction to DFT-LDA, 
but it remains to be seen whether these approaches will be successfully 
applied to a broad range of solid-state materials.

Although there has been much study on the performance of these LDA+U 
functionals in the context of real materials, and an early review of the method
and some applications was provided by Anisimov, Aryasetiawan, and Lichtenstein,\cite{review}
relatively little has been done 
to understand, qualitatively and semi-quantitatively, how the functionals 
operate based solely on their energetics, distinct from DFT-LSDA effects.   
In this paper we analyze the functionals that are commonly used, as well as 
others which were introduced early on but are not so commonly used.  Some of 
the nomenclature in the literature is confusing, so we try to clarify these 
confusions where we can.

\section{The LSDA+U Correction $\Delta E$}

\begin{table*}[th]
\begin{centering}
\vskip 4mm
\begin{tabular}{c|cccc}
LDA+U        & $E_{dc}$ &=& $E_{dc}$    & DFT XC \\
  Functional &          & & (rewritten) &  Functional  \\ \hline \hline
Fl-nS		& $\frac12 UN^2 - \frac{U+2lJ}{2l+1}\frac{1}{4}N^2$				  	   
         &=&  $\frac12 UN^2 - \frac{U+2lJ}{2l+1} \frac{1}{2}\sum_\sigma (\frac{N}{2})^2$ 	& LDA	\\ \hline
Fl-S (AMF) 	& $\frac12 UN^2 - \frac{U +2lJ}{2l+1}\frac{1}{2}\sum_\sigma N_\sigma^2$
         &=&  $\frac12 UN^2 - \frac{U +2lJ}{2l+1}\frac{1}{2}\sum_\sigma N_\sigma^2$     & LSDA	\\     \hline
FLL			& $\frac12 UN(N-1) - \frac12 J\sum_\sigma N_\sigma(N_\sigma-1)$ 
         &=&  $\frac12 UN(N-1) - \frac12 J\sum_\sigma (N_\sigma^2 -N_\sigma)$      & LSDA \\   \hline
FLL-nS		& $\frac12 UN(N-1) - \frac14 J N(N-2)$ 						   
         &=&  $\frac12 UN(N-1) - \frac12 J\sum_\sigma ((\frac{N}{2})^2-N_\sigma)$& LDA	\\     \hline
\end{tabular}
\end{centering}
\caption{The double-counting terms of various LDA+U functionals.  In the second expression two of
them are rewritten to reflect how they are (somewhat deceptively) identical in form, but in one
case a distinction between spin-up and spin-down (relative to half of N: 
$N_{\sigma}\leftrightarrow N$/2) is made.  Note that while the first two forms
appear to contain an isotropic self-interaction 
[$\frac{1}{2}UN^2$ rather than $\frac{1}{2}N(N-1)$] 
they are derived from a form which {\it explicitly} has {\it no} self-interaction between
the orbital fluctuations $\delta_{M\sigma}$. See text for more discussion.}
\label{tbl:Functionals}
\end{table*}
The LDA+U functional is usually coded in a form in which the choice of coordinate system is
irrelevant, often referred to as the rotationally-invariant form.\cite{SashaL95}  
This form involves Coulomb
matrix elements that have four orbital indices, and the orbital occupation numbers are matrices
in orbital space ({\it viz.} $n_{mm'}$).  One can always (after the fact) rotate into the
orbital Hilbert space in which the occupations are diagonal, in which case the interactions
have only two indices.  In our discussion we will work in the diagonal representation. 

The LDA+U functionals considered here can all be written in the form  
\begin{eqnarray}
\Delta E = E_I - E_{dc},
\label{eq:simple}
\end{eqnarray}
 where the direct interaction is
\begin{eqnarray}
E_I = \frac12 {\sum_{m\sigma \neq m'\sigma'}} W_{mm'}^{\sigma\sigma'} n_{m\sigma}n_{m'\sigma'}
\label{eq:LDAU-General}
\end{eqnarray}
and $E_{dc}$ is the double-counting correction.  The Coulomb matrix
elements are given in terms of the direct and (spin-dependent) exchange contributions as
\begin{eqnarray}
W_{mm'}^{\sigma\sigma'} = (U_{mm'} - J_{mm'} \delta_{\sigma,\sigma'}). 
\end{eqnarray}   
By the convention chosen here, $E_I$ and $E_{dc}$ are both positive quantities as long as the
constants $U$ and $J$ (which define the matrix elements $U_{mm'}$ and $J_{mm'}$ but are not the
same) as chosen conventionally, with $U$ much larger than $J$.

Note that the orbital+spin diagonal term has been omitted in Eq. \ref{eq:LDAU-General} -- there is
no self-interaction in $E_I$.  However, it is formally allowed to include the diagonal
`self-interaction' term, because the matrix element vanishes identically (self-interaction
equals self-exchange: $U_{mm} = J_{mm}$), and it can simplify
expressions (sometimes at a cost in clarity) if this is done.  The double counting correction
depends only on the orbital sum $N_{\sigma}$, which appears up to 
quadratic order.  A consequence is that 
it will contain terms in $n_{m\sigma} n_{m\sigma}$,
which are self-interactions.  Thus while the LSDA+U method was not intended as a self-interaction correction
method, it is not totally self-interaction free.  In fact, the underlying LSDA method also contains
self-interaction, and the double-counting term may serve to compensate somewhat this unwanted
effect.  We discuss self-interaction at selected points in this paper.

\subsection{Short formal background to the LSDA+U method.}
The ``LSDA+U method'' is actually a class of functionals.  Each functional
has the same form of interaction $E_I$, with differences specified by \\
(1) choice of the form of double counting term. \\
(2) choice of constants $U$ and $J$. For a given functional, these are
`universal' constants like $\hbar, m, e$, i.e. they are not functional of 
the density in current implementations. 
Possibilities for doing so, that is, determining them self-consistently
within the theory, have been proposed.\cite{erwin} \\
(3) choice of projection method to determine the occupation matrices from the
Kohn-Sham orbitals.  Given identical choices for
(1) and (2) above, there will be some (typically small) differences in
results from different codes due to the
projection method.

The occupation numbers (or, more generally, matrices) are functionals of the
density, $n_{m\sigma}[\rho]$, through their dependence on the Kohn-Sham orbitals.
Then, whereas in LSDA one uses the functional derivative
\begin{eqnarray}
{\rm LSDA}:~~\frac{\partial E_{LSDA}[\{\rho_{s}\}]}{\partial \rho_{\sigma}(r)}
\end{eqnarray}
in minimizing the functional, in LSDA+U the expression generalizes to
\begin{eqnarray}
{\rm LSDA+U}:~~\frac{\partial ( E_{LSDA}[\{\rho_{s}\}] + 
    \Delta E[\{n_{ms}[\rho_s]\}] )}{\partial \rho_{\sigma}(r)}
\end{eqnarray}
Since the resulting spin densities $\rho_{s}$ are changed by including the 
$\Delta E$ correction, the change in energy involves not only $\Delta E$ but also the
change in $E_{LSDA}$.  In practice, there is no reason to compare $E_{LSDA+U}$
with $E_{LSDA}$ as they are such different functionals.  However, in the 
following we will be assessing the importance of the choice of the double
counting term in the LSDA+U functional, and it is of interest
to compare, for fixed $U$ and $J$, the energy differences between 
LSDA+U functionals differing only in their double counting terms in order
to understand the differing results.
Even if the set of occupation numbers turn out to be
the same (a situation we consider below), 
the densities $\rho_{\sigma}$ will be different and the differences in
$E_{LSDA}$ may become important.

As with the non-kinetic energy terms in $E_{LSDA}$, the functional derivatives
of $\Delta E$ lead to potentials in the Kohn-Sham equation.  These are non-local
potentials, which (via the same projection used to define the occupation
numbers) give rise to orbital-dependent (nonlocal) potentials
\begin{eqnarray}
v_{m\sigma} \equiv \frac{\partial \Delta E}{\partial n_{m\sigma}} = 
  v_{m\sigma}^I - v_{m\sigma}^{dc},\nonumber \\
  v_{m\sigma}^I = \sum_{m'\sigma' \neq m\sigma} W_{mm'}^{\sigma\sigma'} n_{m'\sigma'}.
\end{eqnarray}
The double counting orbital potential is discussed later.

The corresponding contribution to the eigenvalue sum $E_{sum}$ is
\begin{eqnarray}
\Delta E_{sum} = \sum_{m\sigma} v_{m\sigma} n_{m\sigma},
\end{eqnarray}
which is subtracted from the eigenvalue sum to obtain the Kohn-Sham
kinetic energy.
However, there are indirect effects of the orbital potentials that affect
all of the kinetic and (LSDA) potential energies; these will be different
for different $\Delta E$ functionals because the orbital potentials, which depend 
on the derivative of $\Delta E$ and not simply on the values of $n_{m\sigma}$,
differ for each functional.  This makes it necessary, for understanding the
effects of the $\Delta E$ correction and the change in energy, to analyze the
orbital potentials. 
We provide a brief discussion in Sec. V.

\subsection{Fluctuation forms of LSDA+U}

First we consider the class of functionals that can be written in what is termed here as a 
{\it fluctuation form}.  The original LDA+U functional was introduced in 1991 by Anisimov, Zaanen 
and Andersen\cite{OriginalLDAU} and was written as
\begin{equation}
\label{eq:Fl-nS}
	\Delta E^{Fl-nS} = \frac12 \sum_{m\sigma \neq m'\sigma'} W_{mm'}^{\sigma\sigma'} 
                            (n_{m\sigma} - \bar n)(n_{m'\sigma'} - \bar n),
\end{equation}
where $\bar n = N^\text{corr} / 2(2l+1)$ is the average occupation of the correlated orbitals.  
(Henceforth $N \equiv N^\text{corr}$.)  Note that the energy is changed only according
to {\it angular `fluctuation'} away from the (spin-independent) angular average occupation.
This form is properly used with LDA (the `LDA averages' $\bar n$ are the reference) and not LSDA. 
This form was  originally advocated with generic  $(U - J \delta_{\sigma,\sigma'})$ matrix
elements instead of the full Coulomb matrix, 
but we use the full $W_{mm'}^{\sigma\sigma'}$ here for comparison with other functionals.  

In 1994, Czyzyk and Sawatzky\cite{CS} introduced a change to (\ref{eq:Fl-nS}) and also
proposed a 
new functional.  The motivation for changing (\ref{eq:Fl-nS}) was to use an LSDA exchange-correlation
functional to treat spin splitting effects rather than LDA.  This change motivated the following equation,
%
\begin{eqnarray}
\label{eq:AMF}
	\Delta E^{Fl-S}&=&\frac12 \sum_{m\sigma \neq m'\sigma'} W_{mm}^{\sigma\sigma'} 
        (n_{m\sigma} - \bar n_\sigma)(n_{m'\sigma'} - \bar n_{\sigma'})\nonumber \\
                   & = & \Delta E^{AMF}
\end{eqnarray}
%
where $\bar n_\sigma = N_\sigma / (2l+1)$ is the average occupation of a single spin 
of the correlated orbitals.  Here the energy correction is due to {\it angular fluctuations}
away from the spin-dependent angular mean, and hence must be used with LSDA.   
We point out that the authors in 
Ref. [\onlinecite{CS}] refer to Eq. (\ref{eq:Fl-nS}) as 
$E^\text{LDA+AMF}$ and Eq. (\ref{eq:AMF}) as $E^\text{LSDA+AMF}$.  This wording may have
caused subsequent confusion, due to the way these terms have come to be used, and also
because a discussion of the ``+U'' functionals requires explicit 
specification of whether LDA or LSDA is being used just to understand which functional 
is being discussed.   Also confusing is that Solovyev, Dederichs, and 
Anisimov\cite{Solovyev94} rejustified Eq. \ref{eq:Fl-nS} using ``atomic limit''
terminology.

The fluctuation forms of LSDA+U are automatically particle-hole symmetric, since
$n_{m\sigma}\rightarrow 1-n_{m\sigma}$, $\bar n_{\sigma}\rightarrow 1-\bar n$
gives $ n_{m\sigma}-\bar n_{\sigma} \rightarrow -(n_{m\sigma}-\bar n_{\sigma})$
and the expression is quadratic in these fluctuations.  The general form of Eq.
\ref{eq:simple} need not be particle-hole symmetric.

Many authors (present authors included) have used 
the term LDA+U where the term 
LSDA+U would be more appropriate, which is especially confusing when discussing the 
AMF functional.  We choose to depart from this confusing nomenclature by giving 
(\ref{eq:Fl-nS}) and (\ref{eq:AMF}) unique names specifying their fluctuation forms, and their connection to LSDA (Fl-S) or to LDA (Fl-nS).
We collect the double-counting terms for the various functionals, along with their
connection to LDA or LSDA, in Table I.

\subsection{The Fully Localized Limit (FLL) Functional}
The second functional introduced by Czyzyk and Sawatzky\cite{CS} is the FLL functional.  
(A $J$=0 version of FLL was
introduced in 1993 by Anisimov {\em et al.}\cite{Anisimov93})  
The authors referred to it (confusingly, as terminology has progressed)
as the ``around mean field'' functional but the atomic 
limit double counting term; in the literature it is now referred to as 
the atomic limit, or fully localized limit (FLL) functional.  This functional cannot be 
written in the fluctuation form 
of the previous two 
functionals (the fluctuation form is exhausted by the -S and -nS cases).   
The FLL functional is written in the form of (\ref{eq:simple}), with 
the double-counting term given in Table \ref{tbl:Functionals}.  

There is yet another LDA+U functional that is available, which was 
introduced in 1993 by Anisimov {\em et al.} \cite{Anisimov93}
There is no clear name for it, but since it can be obtained by using 
$N_\sigma = N/2$ in $E_{dc}$ for FLL, one might consistently refer to 
it as FLL-nS, corresponding to FLL with no 
spin dependence.  The authors in Ref. 
\onlinecite{Anisimov93} indicate that this functional is to be used with LDA,
in accordance with the lack of spin dependence in the double counting term.

\subsection{Implementation of LSDA+U \\
            in Some Widely Used Codes}

The Fl-nS  functional is implemented in the {\sc Wien2k} code, as {\verb nldau } $=2$, and called HMF (Hubbard in Mean Field),\cite{Wien2k} however it is apparently not often used. \\
The Fl-S (AMF) functional is implemented in the {\sc Wien2k} code\cite{Wien2k} as {\verb nldau } $=0$ and the FPLO code\cite{FPLO} as AMF.  It is also available in the {\sc Abinit} code\cite{abinit} when using a PAW basis set\cite{amadon:155104} by setting {\verb usepawu } $=2$.  \\
The FLL functional is implemented in several general-purpose DFT codes, such as {\sc Wien2k} ({\verb nldau } $=1$)\cite{Wien2k}, FPLO (select AL in fedit)\cite{FPLO}, VASP, PW/SCF, and {\sc Abinit} ({\verb usepawu } $=1$) when using a PAW basis set. \\
The FLL-nS functional is available in VASP. \\

\subsection{General Remarks}
When the Fl-nS and Fl-S functionals are written  
in their fluctuation form, there is no 
separate double-counting term, hence one does not need the double-counting interpretation.
They can of course be expanded to 
be written in the `interaction minus double-counting' form of (\ref{eq:LDAU-General}),
which is useful especially for comparing with functionals that can only be written in that form.  
A comparison of the double-counting terms is given in Table \ref{tbl:Functionals}.
Reducing all to interaction minus double-counting form makes the difference between 
the functionals most evident; since they all have the 
same ``direct-interaction'' term, the {\it only difference} between the functionals 
is what double-counting energy is used; the uninteresting tail seems to be wagging the
exciting dog, which is in fact the case.  The double-counting terms can be reduced 
to dependence only on $N$ and $N_\sigma$  thanks to summation rules 
(there is at least one free index) on the 
$U_{mm'}$ and $J_{mm'}$ matrices,
\begin{eqnarray}
	\sum_m U_{mm'} & = &(2l+1)U				\\
	\sum_m J_{mm'} & = &U + (2l)J, 
\end{eqnarray}
that is, the sum over any column (or row) of the U and J matrices is a fixed simple value, which depends on the input parameters $U$ and $J$.  One can then simply see that a sum over a column of $W$ is $(2l+1)U$ if $\sigma \neq \sigma'$ and $2l(U - J)$ if $\sigma = \sigma'$.

The $U_{mm'}$ and $J_{mm'}$ matrices satisfy, by definition, $U_{mm} - J_{mm} = 0$, so that 
there is no self-interaction, whether or not the (vanishing)diagonal term $m\sigma = m'\sigma'$ is included 
in the interaction term.  As mentioned earlier, the following analysis assumes the 
the occupation matrix has been diagonalized.  While this can always be done, 
the transformed matrix elements $U_{mm'}$ and $J_{mm'}$ will not be exactly what we have used in 
Sec. \ref{Sec:NumericalResults}.  

\section{Analysis of the Functionals}

\subsubsection{The $J=0$ simplification.} 
It is not uncommon for practitioners to use `effective' values
${\tilde U} = U-J, {\tilde J}=0$ and insert
these constants (for $U, J$) into LDSA+U.
For $J = 0$, of course Hund's coupling (intra-atomic exchange)
is lost, but $J$ also controls the
anisotropy of the interaction, and for $J=0$ anisotropy also is lost 
($U_{mm'}\equiv U$
as well as $J_{mm'}\equiv 0$ for $m\neq m'$). 
This case is relatively simple, it seems it should provide the ``big picture''
of what LSDA+U does with simple Coulomb repulsion, 
and it has been discussed several times before.
With $J = 0$, the fluctuation functionals simplify to
\begin{eqnarray}
\Delta E^{Fl-\kappa}_{J=0}&=&\frac{U}{2}\sum_{m\sigma \neq m'\sigma'} \delta n_{m\sigma}
                                     \delta n_{m'\sigma'} \nonumber \\
 &=&\frac{U}2 \left[\left(\sum_{m\sigma}\delta n_{m\sigma} \right)^2
             - \sum_{m\sigma}(\delta n_{m\sigma})^2\right] \nonumber \\
 &=&-\frac{U}2 \sum_{m\sigma}(\delta n_{m\sigma})^2 \equiv -\frac{U}2 \Gamma_{\kappa}^2 \leq 0,
\end{eqnarray}
because the sum of fluctuations vanishes by definition for either form 
$\kappa = nS$ or $S$; note the `sign change' of this expression when the diagonal
terms are added, and subtracted, to simplify the expression. 
Here $\Gamma^2$ is the sum of the squares of the fluctuations, bounded
by $0 \leq \Gamma_{\kappa}^2 \leq N$.
For integer occupations the energy corrections for Fl-nS and Fl-S (AMF) can be written
\begin{eqnarray}
\Delta E^{Fl-S}_{J=0}&=&-\frac{U}2 \left[N(1 - \bar n) -\frac{M^2}{2(2l+1)}\right],\nonumber \\
\Delta E^{Fl-nS}_{J=0}&=&-\frac{U}2 N(1 - \bar n).
\label{eq:J=0}
\end{eqnarray}

There are two things to note here.
\begin{enumerate}
\item In Fl-nS, the energy is independent of both the spin and orbital polarization of the state,
which lacks the basic objective of what LSDA+U is intended to model.  
Considering the form of its double counting term (see Table \ref{tbl:Functionals}) 
with its self-interaction
term (proportionality to $N^2$), Fl-nS for $J$=0 becomes simply a self-interaction correction method. 
\item In Fl-S (AMF), configurations with magnetic moments are energetically {\it penalized},
{\it proportionally to} $U$ and quadratically with $M$. 
In later sections we will 
discuss the partial cancellation with the LSDA magnetic energy.	
\end{enumerate}

Under the same conditions, the FLL functional becomes
\begin{eqnarray}
\Delta E^{FLL} = \frac{U}{2}\sum_{m\sigma} n_{m\sigma}(1 - n_{m\sigma}) \geq 0.
\end{eqnarray}
Solovyev {\it et al.}\cite{Solovyev94} noted the important and easily
recognizable characteristics of this expression.  Besides being non-negative,
for integer occupations the energy vanishes.  It is a simple inverted parabola
as a function of each $n_{m\sigma}$.  From the derivative, the orbital potentials
are linear functions of $n_{m\sigma}$, with a discontinuity of $U$ when $n_{m\sigma}$
crosses an integer value.  These characteristics underlie the most basic properties
of the LSDA+U method: integer occupations are energetically preferred, and 
discontinuities in the potentials model realistically the Mott insulator gap that
occurs in strongly interacting systems at (and only at) integer filling.

\subsubsection{$J \neq 0$, but Isotropic}
Simplification of the full expression for a functional results by separating out
the isotropic parts of the interaction:
\begin{subequations}
	\label{eq:UJbar}
	\begin{equation}
		\label{eq:Ubar}
		U_{mm'} = U + \Delta U_{mm'}
	\end{equation}
	\begin{equation}
		\label{eq:Jbar}	
		J_{mm'} =  U\delta_{mm'} + J(1-\delta_{mm'}) + \Delta J_{mm'}.	
	\end{equation}
\end{subequations}
%
%
%
%
%
%
%
%
%
The isotropic parts simplify, giving
\begin{eqnarray}
\Delta E^{Fl-nS}&=&-\frac{U-J}{2}\sum_{m\sigma}n_{m\sigma}^{2}-\frac{J}{4}M^{2}
  \nonumber \\
              & &+\frac{U-J}{2}N\overline{n}+\Delta E^\mathrm{aniso},\\
\Delta E^{Fl-S}&=&-\frac{U-J}{2}\sum_{m\sigma}n_{m\sigma}^{2}
           +\frac{U-J}{4}\frac{M^{2}}{2l+1} \nonumber \\
             & &+ \frac{U-J}{2}N\overline{n}+\Delta E^\mathrm{aniso}, \\
\Delta E^{FLL}&=&-\frac{U-J}{2}\sum_{m\sigma}n_{m\sigma}^{2}+\frac{U-J}{2}N
           \nonumber \\
           &  &+ \Delta E^\mathrm{aniso}
\label{eq:isotropic}
\end{eqnarray}
with the universal anisotropy contribution
\begin{eqnarray}
\Delta E^\mathrm{aniso}&=&\frac{1}{2}\sum_{mm^{\prime}\sigma\sigma^{\prime}}
  \Delta W_{mm^{\prime}}^{\sigma\sigma^{\prime}}n_{m\sigma}n_{m^{\prime}\sigma^{\prime}},\\
\Delta W_{mm'}^{\sigma\sigma'}&\equiv &\Delta U_{mm'} - \Delta J_{mm'} \delta_{\sigma\sigma'}.
\label{eqn:aniso}
\end{eqnarray}
is the anisotropic part of the interaction matrix elements.  These equations, up to the 
$\Delta W$ term, are the extensions of Eq. (\ref{eq:J=0}) to include isotropic exchange in
explicit form.

The first term in each of these expressions contains $-\frac{1}{2}{\tilde U}n_{m\sigma}^{2}
({\tilde U}\equiv U-J)$ and hence has the appearance of a self-interaction correction.
Since the diagonal term of the interaction $E_I$ is specifically excluded, it does not
actually contain any self-interaction; in fact, the sign of the interaction $E_I$ is
{\it positive}.   (The double counting term does contain terms quadratic in $N$ which
must be interpreted as self-interaction.)  Nevertheless, the rewriting of the functional
leads to a self-interaction-like form, and that part of the functional will have an
effect related to what appears in the self-interaction-corrected LDA method, but
by an amount proportional to ${\tilde U}$ rather than a direct Coulomb integral, and depending 
on the difference of $n_{m\sigma}$ from the reference occupation, see Sec. IV.

\subsubsection{Fl-nS}
For Fl-nS, if we are restricted to integer occupations (so $n_{m\sigma}^2 = n_{m\sigma}$), 
then $\Gamma^2$ depends only on $N$, so the first term in $\Delta E^{Fl-nS}$ above 
depends only on $N$.  Then, up to corrections in $\Delta U$ 
and $\Delta J$, the state with the largest total spin moment 
will be favored; this is Hund's first rule.  In fact, even with the $\Delta U$ and 
$\Delta J$ terms, the $-{J}M^2/4$ term is still strongly dominant.  Except for $N = 7$, 
there are many ways to arrange electrons in orbitals which maximizes $S$.  Energy differences 
between these arrangements arise only from anisotropy ($\Delta U$ and $\Delta J$) 
and spin-orbit coupling.

\subsubsection{Fl-S}
In Fl-S, instead of having the $-{J}M^2/4$ term from Fl-nS which {\em favors} magnetism, 
there is a term $\frac{(U - J)}{4(2l+1)}M^2$ which {\em opposes} magnetism.  
This term (as in the $J=0$ case) comes from the occupation variance which wants to evenly 
distribute electrons across both spin channels.  Within LSDA there is something like a 
Stoner term of the form $-\frac14 I M^2$ which will compete with this Fl-S magnetic 
penalty.  We return to this aspect in later sections and the appendix.

\subsubsection{Spin-orbit Coupling; Particle-Hole Symmetry}
Without spin-orbit interaction, for a given $N$ there are many states that are 
degenerate for both double counting schemes.  Every value of $N$ has at least four 
degeneracies, those with $\pm L_z, \pm S_z$. 

Any state which has the same number of spin up as spin down electrons ($M=0$) gives the 
same energy from Fl-nS and Fl-S, since then $\bar n_\uparrow = \bar n_\downarrow 
= \bar n$ (the orbital potentials are distinct, however).  
Of course this fixed $N$, $M$=0 specification may contain many different
configurations.  Looking at results mentioned later, for Fl-S the ground state 
for an even number of electrons is $S_z = 0$ (so $\bar n_\sigma = \bar n$), thus the 
configuration which gives the Fl-S ground state has the same energy in Fl-S and Fl-nS.

\section{Fractional Occupations}

Here we briefly discuss the effect of non-integer occupations in LSDA+U.  Taking a 
general set of occupations as $\{n_{m\sigma}\}$, we define a set of integer 
occupations, $\{\hat{n}_{m\sigma}\}$, and the fractional part of the occupations 
as $\gamma_{m\sigma} = n_{m\sigma} - \hat{n}_{m\sigma}$.  For illustration purposes 
we will choose the simplest possible scenario, where charge is transfered to an 
empty orbital $a$ from an occupied orbital $b$ both of the same spin, so that 
$0 < \gamma_{a\uparrow} = -\gamma_{b\uparrow}$, $\hat{n}_{a\uparrow} = 0$ and 
$\hat{n}_{b\uparrow} = 1$.  With this selection, $N_\sigma$ is unchanged 
(and therefore, $N$ and $M$ as well) so that $E_{dc}$ is unchanged.  Thus, the 
effect of the charge transfer is entirely contained in the $E_I$ term.  
Expanding $E_I$ for the general occupation set gives
\begin{equation}
		E_I[\{n_{m\sigma}\}] - E_I[\{\hat{n}_{m\sigma}\}] =
			U\gamma_{a\uparrow} (1 - \gamma_{a\uparrow})
\end{equation}
for the $J = 0$ case, and for $J \neq 0$ we find
\begin{eqnarray}
E_I[\{n_{m\sigma}\}]&-&E_I[\{\hat{n}_{m\sigma}\}]  \nonumber \\
  &=&
   \sum_{m\sigma} (W_{am}^{\uparrow\sigma} - W_{bm}^{\uparrow\sigma}) 
      \hat{n}_{m\sigma} \gamma_{a\uparrow} \nonumber \\
  & &            - W_{ab}^{\uparrow\uparrow} \gamma_{a\uparrow}^2.
		\label{eq:FracOccExp}
\end{eqnarray}

The dominant term in (\ref{eq:FracOccExp}) is where $m\sigma = b\uparrow$.  
This term gives a contribution $W_{ab}^{\uparrow\sigma'}\gamma_{a\uparrow} 
\sim U\gamma_{a\uparrow}$ (since $U >> J$ for typical parameter choices, 
where other terms give contributions proportional to 
$(W_{am}^{\uparrow\sigma} - W_{bm}^{\uparrow\sigma})\gamma_{a\uparrow} \propto 
J\gamma_{a\uparrow}$.  The term with $m\sigma = a\uparrow$ is killed off by 
the factor of $\hat{n}_{a\uparrow}$, and the term in $\gamma^2$ is significantly 
smaller than the others for $\gamma < 0.5$.

This shows that there is an energy penalty for fractional occupation, 
proportional to $U$ and {\em linear in $\gamma$} at small $\gamma$.  Thus, in 
configuration space, the LSDA+U functionals have many local minima around 
configurations with integer occupations.  This result is fairly general.  
Even for charge transfer between orbitals of opposite spins, the linear energy 
penalty in $\gamma$ is still dominant over any additional terms coming from 
the double-counting or spin-orbit.

In practice, this gives the possibility that LSDA+U will get `stuck' in a 
local minimum with some configuration that may not be the true ground state.  
This behavior is not uncommon; LSDA+U has been reported\cite{CeLDAU} to find multiple 
local minima depending on the starting configuration.

\section{Orbital Potential Matrix Elements}
Up to now only the energy functionals themselves were discussed. 
Now we return to the derivatives, the orbital potentials $v_{m\sigma}$.
It is simple to derive the exact expressions, and the interaction term $E_I$
common to all forms gives a potential $\Delta v_{m\sigma}$ which depends only
on the occupations of the {\it other} orbitals 
 $n_{m'\sigma'}, m'\sigma' \neq m\sigma$.
The potential resulting from the double counting term is functional specific, and
may contain a contribution from $n_{m\sigma}$ itself, {\i.e.} a self-interaction.

We confine our observations here to the subdivision (introduced just above) 
of the interaction into
an unitarily invariant isotropic part, and into an anisotropic part Eq.
(\ref{eq:UJbar}) that is much smaller and more difficult
to analyze.  As for the energy itself, it is convenient to add and subtract the
diagonal self-Coulomb and self-exchange, which makes the effect of the potential
much more transparent at the cost of introducing the misleading self-interaction
interpretation.

The potential matrix elements are
\begin{eqnarray}
\Delta v_{m\sigma}^{Fl-nS}&=&-\left(U-J\right)\left[n_{m\sigma} -\overline{n}\right]
         -\frac{J}{2}M\sigma \nonumber \\
                          & &+\Delta v_{m\sigma}^{aniso},\\
\Delta v_{m\sigma}^{Fl-S}&=&-\left(U-J\right)\left[n_{m\sigma} - \overline{n}_{\sigma}\right]
         +\frac{U-J}{2}\frac{M}{2l+1}\sigma \nonumber \\
                         & &+\Delta v_{m\sigma}^{aniso},\\
\Delta v_{m\sigma}^{FLL}&=&-\left(U-J\right)\left[n_{m\sigma} - \frac{1}{2}\right]
          +\Delta v_{m\sigma}^\mathrm{aniso},
\end{eqnarray}
with the anisotropic potential term 
\begin{eqnarray}
\Delta v_{m\sigma}^\mathrm{aniso}=\sum_{m^{\prime}\sigma^{\prime}}
         \Delta W_{mm^{\prime}}^{\sigma\sigma^{\prime}}
                n_{m^{\prime}\sigma^{\prime}}.
\end{eqnarray}
The main occupation number dependent term, proportional to $n_{m\sigma}$,
has a self-interaction appearance and effect, as discussed above for the
functionals.  The differences in this term arise from the ``reference'' occupation
with which $n_{m\sigma}$ is compared to determine the potential shift.  The 
``fluctuation'' $n_{m\sigma} - n_\mathrm{ref}$ is smaller for $Fl-S$ (AMF) than for $Fl-nS$
because the occupation for a given spin direction tends to be closer to $\bar{n}_{\sigma}$
than to $\bar{n}$.  The reference occupation for FLL is, like $Fl-nS$, spin-independent,
in fact, the reference is half-filling. In this sense, FLL seems more like a single-band
Hubbard model treatment than the other two functionals.

The other difference that is evident in this form is the spin dependence.
$Fl-nS$ additionally has a spin orientation dependent potential shift 
proportional to $J$ and to $M$ (similar to an LSDA treatment, but using $J$ instead of
the Stoner $I$) and enhances spin-splitting 
of the eigenenergies $\varepsilon$ accordingly.  
In Fl-S (AMF) the analogous term is $+\left(U-J\right)\frac{M}{2\left(2l+1\right)}\sigma$,
with a sign that impedes magnetism.
It can be simplified to  $\approx \frac{J}{2}M\sigma$
when $U\approx2\left(l+1\right)J$ . This expression illuminates the
reason that AFM is sometimes found to {\it decrease} the magnetic moment:
this term more or less cancels the spin
splitting of LSDA due to the opposite sign. What is left is a splitting
of occupied and unoccupied levels due to the $n_{m\sigma}$ term, which is almost
independent of $M$.  The effect is to support a spin-polarized solution,
but provide little discrimination between different $M$. Since the
spin polarization energy does not favor large $M$, we end up with
a tendency of a near degeneracy of different $M$ values,
as we already pointed out from purely energetic arguments.
For the case of a half-filled fully polarized shell $n_{ms}=\delta_{\sigma,1}$
(the case $N=M=7$ in Section \ref{Sec:NumericalResults}) the potential
matrix vanishes, which can be seen from $\overline{n}=\frac{1}{2}$
$\frac{M}{2\left(2l+1\right)}=\overline{n}=\frac{1}{2}$. However,
at the same time the energy contribution also vanishes $\Delta E^{Fl-S}=0$
(for integer occupations) and the Fl-S functional has no effect at
all.

The SIC term in FLL splits occupied
and unoccupied states symmetrically, while in the fluctuation functionals
the splitting happens with respect to the averaged occupation, which
is seen in the overall energy positions in Fig \ref{fig:Scatter}.


\section{Numerical Results}
\label{Sec:NumericalResults}
\begin{figure}[t]
\includegraphics[width=0.45\textwidth]{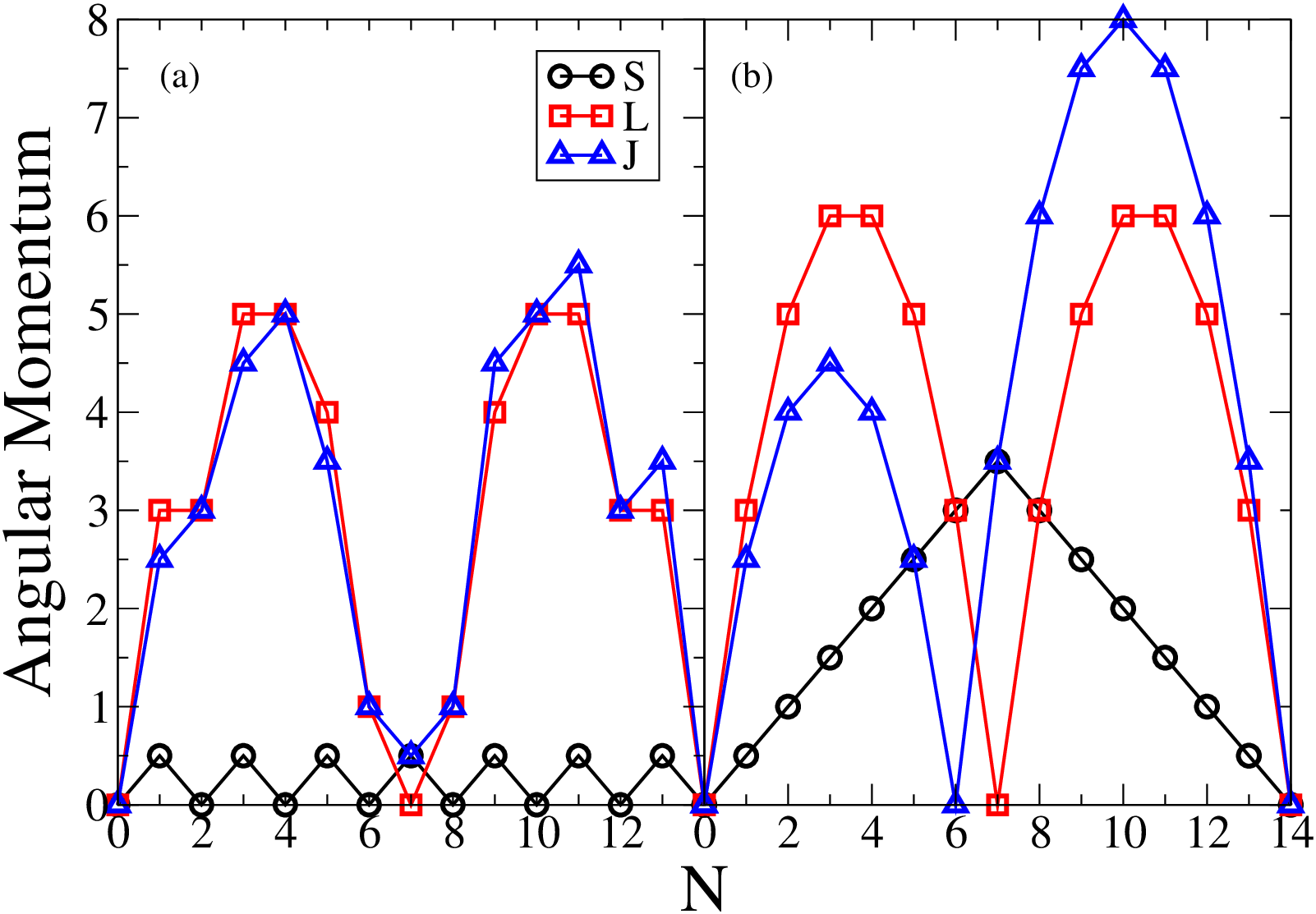}
\caption{(Color online) Angular momentum values of $S_z, L_z$, and $J_z$ 
of the lowest energy state for (a) AMF (Fl-S) and (b) FLL, with 
spin-orbit coupling.  Parameter values are $U = 8$, $J = 1$, $I = 0.75$.  
The AMF (Fl-S) curves do not follow 
Hund's rules, because the Stoner parameter is too small.  FLL follows 
Hund's rules exactly with these parameters.
}
\label{fig:AngSO}
\end{figure}

Following common terminology, for the remainder of the paper we refer to the
Fl-S functional simply as the AMF form.
We have taken values for $U_{mm'}$ and $J_{mm'}$ (used for Eu) from 
Ref. \onlinecite{EuN} (recalculated, to include more significant figures).  
These matrices are generated using 
$U = 8$ and $J = 1$ (values typical of rare earths) following the 
procedure given in the appendix of Ref. \onlinecite{CS}.

In our analysis of the AMF and FLL functionals, which are based on
an LSDA reference state, we include a Stoner term
\begin{eqnarray}
	E(M) = -\frac14 I M^2
	\label{eq:Stoner}
\end{eqnarray}
to model the magnetic effects of LSDA on the total energy.  The 
addition of this term helps to give a picture of the degree to which the
functionals reproduce Hund's first rule.  
Typical values of $I$ for ionized lanthanides are 0.75 eV, so we use 
this value for the calculations of this section.  Further discussion 
of the Stoner $I$ is included in the appendix.  

Spin-orbit interaction is included in the form
\begin{equation}
	E_{SO} = \lambda \vec S \cdot \vec L \rightarrow
                \sum_{m\sigma} S_z  L_z,
\end{equation}
where the second form applies when only $z$-components of moments
are treated, as is done in current implementations of the LSDA+U method.
Due to this restriction, LSDA+U often does not produce the correct multiplet 
energies in the atomic limit.  The visible result in LSDA+U band structures
is splittings of occupied, or unoccupied, correlated suborbitals that
can be as large as a few times $J$, and understanding the splittings is
not straightforward.  For $4f$ systems these splittings\cite{EuN,ReN}
may not be of much interest unless one of the correlated bands approaches
the Fermi level. In heavy fermion compounds, for example, LSDA+U results are
used to infer which parts of the Fermi surface has a larger amount of 
$f$ character.\cite{YbRh2Si2}  The same effects (eigenvalue splittings)
occur in $3d$ or $5f$ systems,
however, where they are expected to become more relevant but are 
masked by stronger banding tendencies.

\begin{figure}[t]
\includegraphics[width=0.45\textwidth]{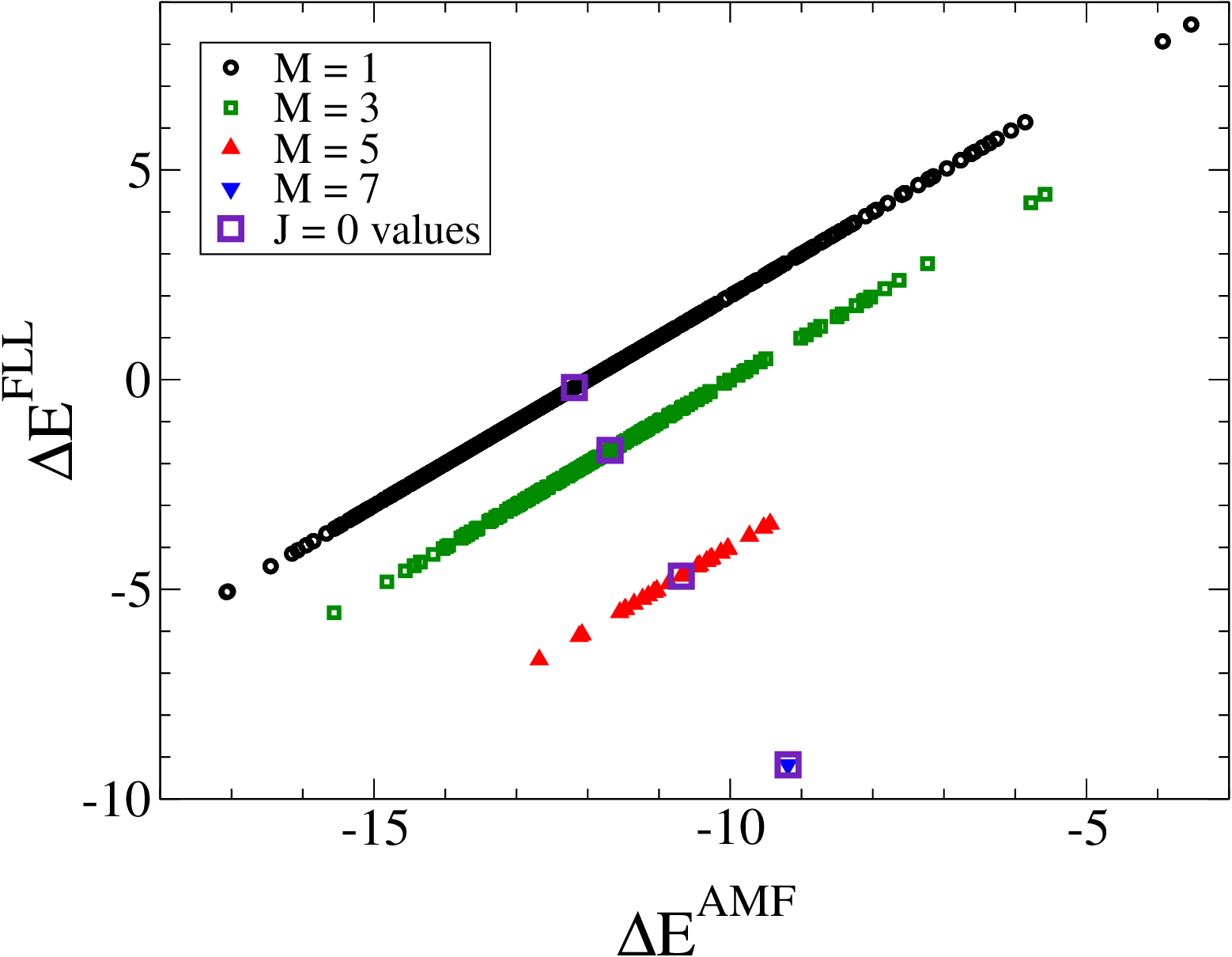}
\caption{(Color online) Shown here is $\Delta E^{FLL}$ plotted vs. 
$\Delta E^{Fl-S}$ for each of the 3432 configurations of $N=7$ electrons, using 
$U=8, J=1, I = 0.75$, all in eV.  The ordering of states is shown for Fl-S by 
counting from left to right, and for FLL by counting from bottom to top.  
Open squares show values for $U = 7$ and $J = 0$.     }
\label{fig:FLL-AMF}
\end{figure}

Here we consider values of $\lambda$ of 0 and 0.2 eV.  The magnitude 
of the spin-orbit interaction is not critical to the results; it mainly 
serves to break degeneracies.  Without the spin-orbit interaction, the 
ground state for any of the functionals at a given $N$ is degenerate 
with several other states.  For instance with $N=6$, the AMF functional
has states 
with $L_z=1, S_z=0$ and $L=11, S_z=0$ with the same lowest energy.

In Fig. \ref{fig:AngSO} are the ground states for both AMF and FLL with 
$U = 8$, $J = 1$ and $I = 0.75$.  The FLL and Fl-nS (not shown) schemes 
both reproduce Hund's rules exactly with these parameters.  AMF does not 
reproduce Hund's rules (in fact penalizes magnetism)
until $I$ is increased to around 1.5, which is 
somewhat larger than reasonable values of $I$.  If one expects LSDA+U to 
reproduce Hund's rules, then the AMF scheme performs rather poorly.  
For instance, at $N = 7$, Hund's rules ask that all electrons be 
spin-aligned, but the AMF ground state has only one unpaired spin  
due to the magnetic penalty appearing in Eq.
(\ref{eq:J=0}).  With these parameter choices, $U/(2l+1) > I$, so 
the AMF magnetic penalty wins over the Stoner energy.  This is likely to 
be the case for $3d$ transition metals as well, since $U_{3d}/(2l+1) 
\sim 1$eV, but it may not be as significant since $I$ for $3d$ elements 
is larger.

We examine the energetics in more detail in Fig. \ref{fig:FLL-AMF}, where 
$\Delta E$ for the AMF and FLL functional is plotted for every 
configuration for $N = 7$.  
The configurations fall into separate lines for each spin moment $M$, since
$E_{dc}$ depends only on $N$ and $M$ for both functionals.  
For the case of $J = 0$, all the states with a particular $M$ value 
collapse to a single energy value (the orbital index loses any impact), 
this is shown with the open squares.  
A value of $I$ was chosen so that the cancellation discussed in the 
previous paragraph is slightly broken.  

If we examine the $J = 0$ case first (the large open squares in Fig. \ref{fig:FLL-AMF}),
we see that the separation of states in FLL is much larger than AMF
(9 eV versus 3 eV), with $M = 7$ the lowest energy for FLL but highest for AMF.
This is a direct consequence of the magnetic penalty of AMF discussed previously.  
If $I$ were increased above 1 eV (keeping the other parameters fixed), then AMF would begin to 
favor the $M=7$ state by a small amount.  

Once $J$ is turned on, the degeneracy is split, and the 
configurations with a particular $M$ spread out around the $J = 0$ value.  The spread is especially large
for the highly degenerate $M=1$ value (from -5 to 8 eV), so that even if $I$ were larger than the typical LSDA value 
(in which case, with $J=0$ AMF would favor a high spin state) the large spread of $M=1$ values
would cause the low-spin states to be favored in AMF.  This spread is entirely coming from the $E_I$ term and is
independent of the double-counting choice.
Here we see for AMF a competition between $J$ and $I$:  $J$ is actually preferring 
a low spin configuration, in contrast to the conventional wisdom that $J$ 
increases the tendency for magnetism.  We see this same tendency occurs in FLL, as for $J = 0$ the separation
between $M=7$ and $M=1$ states is 9 eV, but with $J = 1$ this separation is reduced to 4 eV.  
Since in FLL the Hubbard $U$ does not penalize magnetic states the way AMF 
does, the presence of $J$ is not able to compete with $I$.  
This makes it clear why FLL is generally accepted to perform better for systems known 
to have high-spin states ({\it e.g.} Eu and Gd).  Conversely, FLL may be 
less successful at modeling low-spin states.  

As mentioned previously, it is fairly common for theoretical studies to replace $U$ and $J$ with effective parameters $\tilde U$ and $\tilde J$.  
For any double-counting term chosen, using these effective parameters will lower the energy of
the high-spin state relative to the low spin state as compared to using $U$ and $J$ directly.  With orbitals that are not highly localized,
such as $3d$ or $5f$ states it may be the case with FLL that the 
reduction of the energy separation between high-spin and low-spin caused by using $U$ and $J$ 
would allow for significant competition between magnetism and kinetic energy in LSDA+U.

We now have seen why and  how FLL and AMF perform differently in assigning a 
magnetic moment.  This may be of particular interest for studies of 
pressure-induced changes in magnetic moment, such as that seen in 
MnO\cite{Kasinathan07} without changes in orbital $M$ occupancy.  Applications of LSDA+U are more thoroughly discussed 
in Sec. \ref{sec:Applications}.


%
\begin{figure*}[t]
\includegraphics[width=0.95\textwidth]{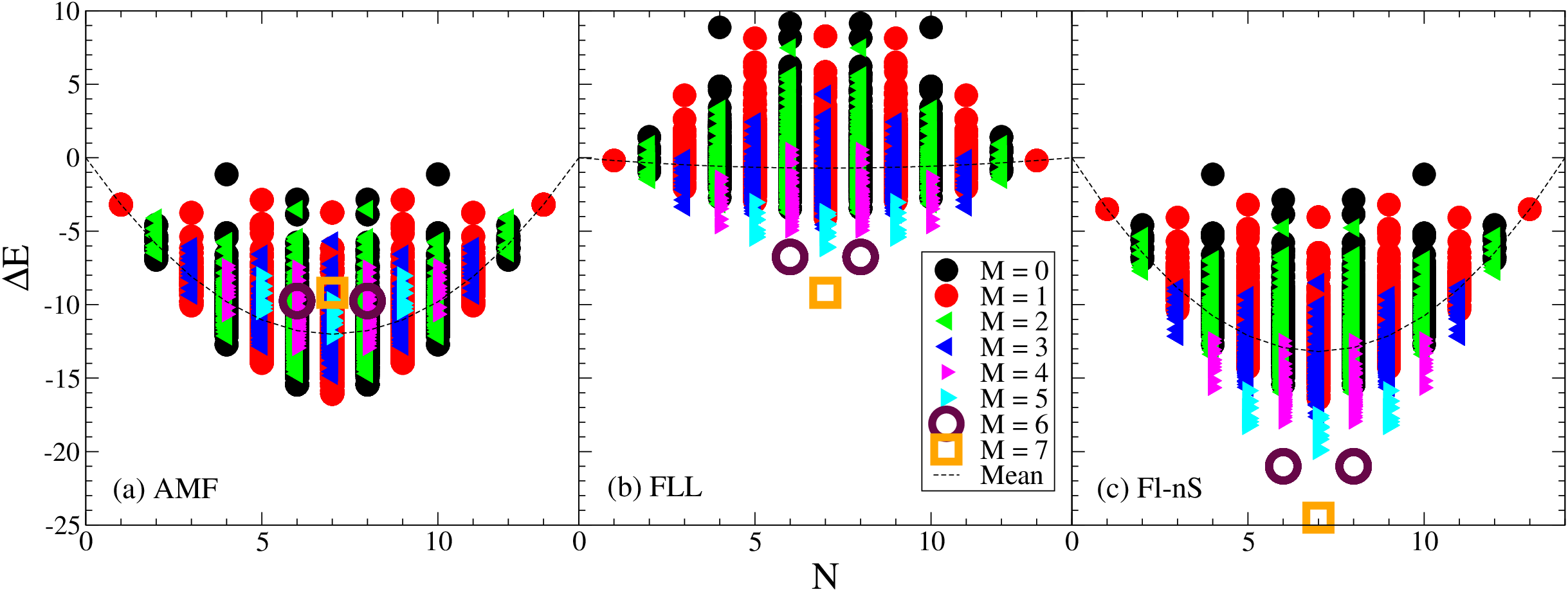}
\caption{(Color online) Scatter plot of all energies $\Delta E$ for 
all states in the (a) AMF(Fl-S), (b) FLL and (c) Fl-nS double-counting 
schemes, for $U = 8$, $J = 1$, and $I = 0.75$ (FLL and AMF only).  
Spin-orbit is neglected here.  For AMF, low spin states (black and 
red circles) appear as lowest energy configurations for all $N$, 
but this is not the case for FLL or Fl-nS.  The dashed lines indicate the
mean energy over configuration for each $N$; note that the variation with
$N$ is much less for FLL than for the other two functionals.
}
\label{fig:Scatter}
\end{figure*}

Shown in Fig. \ref{fig:Scatter} are scatter plots of the energies of all 
possible states for a given number of $f$ electrons with integer 
occupations.  SO is neglected, as it makes very minor changes to this 
picture by splitting some degeneracies.  The particle-hole symmetry of
each functional is apparent.
In Fl-nS and FLL, the ground state energy for 
$N = 7$ is roughly 3 eV lower than the next level, which are the 
(degenerate) ground states 
for $N = 6$ and $8$.  This is almost entirely due to the term depending on 
$M^2$ (either the $J$ term in (\ref{eq:Fl-nS}) or the Stoner term 
in FLL), because $M$ is large and at its maximum with 7 spins aligned.  
In AMF low spin states can be seen at the low end of the range 
for configurations at each $N$; the high spin states for $N$=6 and 7 are 
disfavored by 6-7 eV.  We see that the trend where AMF favors low-spin configurations
and FLL favors high-spin configurations shown for $N=7$ in 
Fig. \ref{fig:FLL-AMF} is present for all $N$.  The large spread of values for low spin configurations (black and red circles)
is seen clearly for AMF as they appear in both the lowest energy positions and the highest energy positions.  The high-spin configurations (large open symbols and triangles) 
are in the middle of each distribution for $N$.  For e.g. $N=5$, counting from the lowest energy, $M=1$ configurations are found first, followed by $M=3$ configurations then the $M=5$ configurations are found (with the trend reversing counting up to the highest energy states).  In FLL, the lowest energy configurations for $N \neq 7$ are still the configurations with maximum spin for a given $N$, and states with lower spins are found in succession.  Again using $N=5$ as an example, the $M=5$ configurations are lowest in energy, and then $M=3$ configurations are seen at energies lower then $M=1$ states.

\section{Discussion}

In this paper we have tried to clarify the behavior of the various functionals that are used in the 
LSDA+U method, we have compared the functionals formally in certain limits, 
we have presented the orbital potentials that arise, and we have
analyzed the total energy corrections that LSDA+U functionals apply 
to LSDA total energies, given a set of occupation numbers.  
The Fl-nS functional which was originally introduced 
strongly favors spin-polarized states as does the commonly used FLL functional. 
The other most commonly used functional besides FLL, Fl-S (AMF), has characteristics that
tend to {\it suppress} moment formation or reduce the magnitude of the moment.  
When analyzed, this AMF functional shows positive energy penalties to magnetism that compete 
with the magnetic tendencies of the LSDA functional, and when $J > 0$ non-magnetic 
solutions become even more
likely to win out.  We have provided a short analysis of the behavior when $J=0$ is used.  While this case is instructive, we advise against its use; it is just as simple to do the full $J \neq 0$ calculation.

When LSDA+U is applied to correlated insulators in the strong coupling regime, it provides
a very good picture of the system at the band structure (effective one-electron) level.
The initial successes include the $3d$ transition metal monoxides MnO, FeO, CoO, and NiO,
for which the LSDA description is very poor.  Other early successes included the insulating
phases of the layered cuprates that become high temperature superconductors when doped, and
the unusual magnetic insulator KCuF$_3$, which was the first case where crucial orbital
ordering was reproduced.  LSDA+U is not a satisfactory theory of
single particle excitations of such systems, but nevertheless provides a realistic picture
of the underlying electronic structure.

The more interesting, and more difficult, cases now lie between the strongly correlated limit of wide-gap 
magnetic insulators and weakly correlated regime that is well described by LSDA.  Some
of these are metals, some are unconventional insulators, and many lie near the metal-insulator
borderline.  It is for these intermediate cases that it becomes essential, if applying the
LSDA+U approach, to understand what the method is likely to do, and especially to understand
the tendencies of the various choices of functional.  This is what we have tried to clarify
in this paper.  As a summary, we will provide an overview of an assortment of results that have 
appeared in the literature for systems that lie somewhere in the intermediate correlation
regime.

\subsection{Examples of LSDA+U behavior from applications}
\label{sec:Applications}
\subsubsection{Strongly correlated insulators.}
{\it Cuprates.} The insulating phase of the cuprate class of high temperature
superconductors comprised the ``killer app'' that served to 
popularize\cite{OriginalLDAU,VIA1992} the LSDA+U method,
and in the intervening years the method has been applied to cuprates and other correlated insulators too many times to cite.  
Simply put, in cuprates it produces the 
Cu $d^9$ ion and accompanying insulating band structure.\cite{VIA1992,rosner}  
The hole resides in the
$d_{x^2-y^2}$ orbital and is strongly hybridized with the planar oxygen $p_{\sigma}$
orbitals, as much experimental data was indicating.

{\it MnO.}
Experimentally, MnO shows at room temperature a moment collapse 
from $M=5$ to $M=1$ (or less), a volume collapse,
and an insulator-to-metal transition, near 100 GPa; this is the classic Mott transition.
Within LSDA, the moment decreases continuously with decreasing volume,\cite{Cohen1997} 
from the high spin (HS) state 
to a low spin (LS) state. The insulator-to-metal transition occurs at much too low a
pressure (without any other change). A volume collapse is predicted, although 
the pressure is significantly overestimated (150 GPa).

The application of LSDA+U in its FLL flavor has been applied and analyzed in
detail,\cite{Kasinathan07} and provides a different picture in several ways. The ambient
pressure band gap is improved compared to experiment. The volume collapse transition 
occurs around 120 GPa and is accompanied by a moment collapse from $M=5$ to $M=1$. 
The nature of this (zero temperature) transition is insulator to insulator, while the
experimental data indicate an insulator-to-metal transition at room temperature. 
The zero temperature transition might indeed be insulator-to-insulator;
such a phase transition would be a type that LSDA+U should work well for.
It is also possible that the static mean-field approximation underlying LSDA+U, which favors
integer occupations and hence insulating solutions, has too strong a tendency and fails
to describe this transition.  This question could be settled by studying experimentally
the Mott transition at low temperature

Even more unexpected than the insulator to insulator aspect is the LSDA+U 
prediction is that the low spin state has 
an unanticipated orbital occupation pattern,\cite{Kasinathan07} being one
in which every $3d$ orbital remains singly occupied (as in the high spin state).
but spin in two orbitals antialign with those in the other three orbitals.
This state is obtained simply from the $M=5$ high spin state by
flipping the spins of two of the orbitals. The resulting density remains spherical, but
the spin density exhibits an angular nodal structure leading at the same time
to a high degree of polarization of the spin-density but a low total moment ($M=1$).
This solution (being the high pressure ground state in LSDA+U) can be traced\cite{Kasinathan07} back
to the interplay between symmetry lowering due to the antiferromagnetic order
(cubic lowered to rhombohedral) 
and the anisotropy part of the interaction Eq. \ref{eqn:aniso}. The symmetry lowering
lifts the cubic grouping ($t_{2g}$ and $e_g$ manifolds), thus allowing a higher number
of allowed occupation patterns.

The anisotropic part of the interaction is responsible\cite{Kasinathan07} 
for Hund's second rule ordering
of states, which has the tendency to increase the mutual distance of each pair
of electrons. If the overall energetics (band broadening and kinetic effects) 
reduce the gain of energy due to spin-polarization, then Hund's 
first rule may become suppressed and the result is a low spin state.
The anisotropic interaction is however not influenced
by this suppression, since it is a local term proportional to a parameter $J$.
It will enforce a Hund's second rule like separation of the electrons under the
low spin condition, and thus can be shown to result exactly in the occupation pattern observed
for MnO. In a sense the low spin state is an example of Hund's second rule without 
Hund's first rule.

{\it FeO, CoO, NiO.} Together with MnO, these classic Mott (or `charge transfer')
insulators have been prime applications of the LSDA+U 
method.\cite{AnisimovNiO94,ShickLDAU,Bengone00,Dobysheva04}  The behavior of the
open $3d$ shell in these compounds has not been analyzed in the detail that was
done for MnO, however.

\subsubsection{Metals}
Correlated metals involve carriers that can move, hence they invariably involve
fluctuations, in occupation number, in magnetic moment, in orbital occupation, etc.
It cannot be expected that a self-consistent mean field treatment such as LSDA+U can
answer many of the questions raised by their behavior.  However, there is still 
the question of whether LSDA+U can provide a more
reasonable starting point than LSDA alone in understanding these metals.  
In our opinion, this remains 
an open question, but one for which some evidence is available.

The Fe-Al system has provided one platform for the application of LSDA+U to moderately
correlated metals.  The systems treated include the Fe impurity in Al (Kondo system,
experimentally), and the compounds Fe$_3$Al, FeAl, and FeAl$_3$. The calculated behavior is too 
complex to summarize here.  The LSDA+U result will, generally speaking, be likely to
give a good picture of a Kondo ion when it produces an integer-valent ion with a
large value of U.  Both FLL and AMF
functionals have been applied in this regime,\cite{Mohn01,Lechermann04} with substantially differing
results, leading one to question whether either is more realistic than simple LSDA.  
Results are also sensitive to volume, {\it i.e.} whether using the experimental
lattice constant or the calculated equilibrium value, and the calculated equilibrium
is different from LSDA and LSDA+U. One result was that, for moderate 
U$_{Fe} \sim$ 3-4 eV,  AMF strongly 
reduces the magnetic moment, while FLL does not.\cite{Lechermann04}
Another application found that the magnetism disappeared within a certain range of
intermediate values of $U_{Fe}$, that is, it was magnetic around small $U_{Fe}$ and
also again at large coupling,\cite{Mohn01} but non-magnetic between.


\subsubsection{Moderately strongly interacting oxides.} 
Trying to address seriously the electronic
structure of intermediate coupling oxides, which are often near the metal-insulator
transition, is a challenge that has begun to be addressed more directly.  The peculiar
Na$_x$CoO$_2$ system, which becomes superconducting when hydrated (water intercalates 
between CoO$_2$ layers) is one example.  One set of studies showed no appreciable 
difference between FLL and AMF,\cite{KwanWoo04} with both predicting charge disproportionation on the 
Co ion for $x$=$\frac{1}{3}$ and $\frac{1}{2}$ for $U\approx$ 2.5-3 eV.   It is likely
that this compound presents a case where the interplay between LSDA and $U$ has effects
that are not fully understood.  Also, it is unclear why there is so little difference
between the FLL and AMF functionals in this system.
 
The compound Sr$_2$CoO$_4$ is another example.  Both functionals show a collapse of the 
moment\cite{KwanWoo06} around $U$ = 2.5 eV, related to the metal-half metal transition that occurs,
but the result for the moments ($M$(AMF) $<$ $M$(FLL)) bears out the tendency of AMF to
penalize magnetic moments..  
The fixed spin-moment calculation in Fig. 9 in Ref. \onlinecite{KwanWoo06} is instructive too, showing the competition 
between LSDA magnetic energy and AMF magnetic penalty.  Also it shows the creation of 
local minima around M = integer values that LDA+U introduces.

\subsubsection{f electron materials.}  
{\it 4$f$ systems.}
These metals often display the correlated electron physics of
a magnetic insulator at the band structure level: background conduction bands provide the
metallic nature, while the correlated states have integer occupation.  The LSDA+U method
seems to be a  realistic method for placing the $f$ states closer to where they
belong (away from the Fermi level).  Gd is a good example, which has been studied at
ambient pressure and compared to photoemission data\cite{ShickLDAU} and magnetic dichroism
data.\cite{HarmonGd,Alouani07}  The LSDA+U method has also been applied up to extremely high pressure
to assess where the `Mott transition' in the $4f$ bands is likely to occur.
The LSDA+U method has also been applied to heavy fermion metals, for example Cu and U
compounds,\cite{Oppeneer97} PrOs$_2$Sb$_{12}$,[\onlinecite{Harima05}] and  
YbRh$_2$Si$_2$~[\onlinecite{YinYb}].  In such systems the LSDA+U method may even 
provide a good estimate of which
itinerant states at the Fermi level are strongly coupled to the localized $f$ states, {\it i.e.}
the Kondo coupling matrix elements.  These
$4f$ systems may become heavy fermion metals (YbRh$_2$Si$_2$) or novel heavy fermion
superconductors (YbAlB$_4$), or they may remain magnetic but
otherwise rather uninteresting metals (Gd). 

{\it $5f$ systems.} 
A variety of application of the LSDA+U method to $5f$ systems, and especially Pu, have
been presented.\cite{Bouchet00,Savrasov00,Price00,Shick05,Shorikov05}  
Given the complexity of the phase diagram of elemental Pu, together
with claims that dynamic correlation effects must be included for any realistic
description of Pu, a more critical study of Pu would be useful.





\section{Acknowledgments}
We have benefited from discussion on various aspects of this work with
M. Johannes, J. Kune\v{s}, A. K. McMahan, I. I. Mazin, and G. Sawatzky.
This project was supported by DOE through the Scientific Discovery through 
Advanced Computing (grant DE-FC02-06ER25794), by DOE grant DE-FG02-04ER46111, and by the Computational Materials Science Network.

\appendix
\section{Calculation of the Stoner $I$ for $3d$ and $4f$ Shells}
\label{Appendix}

\begin{figure}
\begin{centering}
\includegraphics[width=0.30\textwidth]{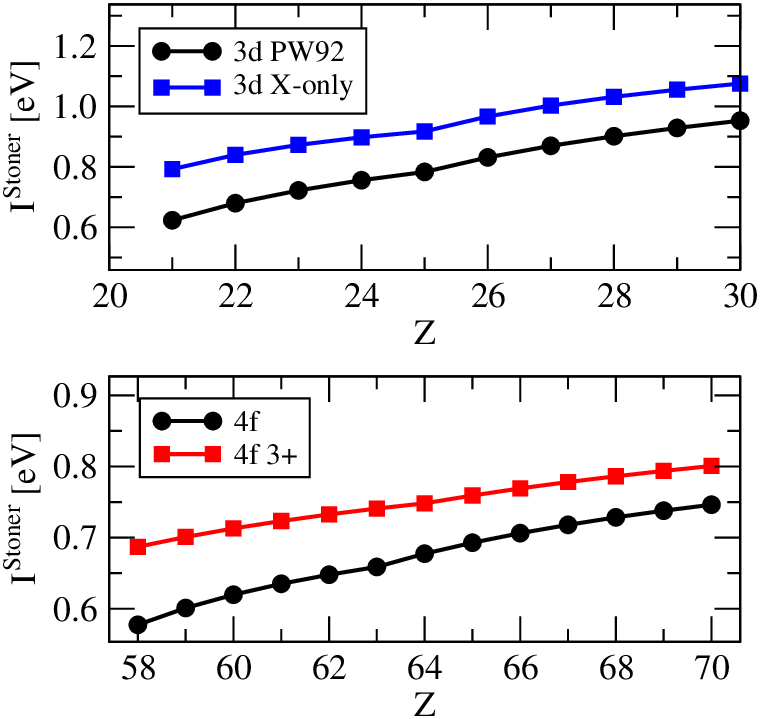}
\par\end{centering}

\caption{\label{fig:Inl}Shell-Stoner integrals for the $3d$ and $4f$ atoms.
For explanations see text.}

\end{figure}

The Stoner parameter $I$ is a well established quantity. For 
metals its value is obtained by a second order expansion
of the LSDA xc-energy around the non-magnetic solution, resulting
in a Fermi surface averaged integral of the radial wave functions
with the xc-kernel.\cite{Janak77} LSDA+U is usually
applied to describe insulating states, where the Fermi surface vanishes.
In the context of discussing the LSDA contribution to the
energy of a correlated $d$- or $f$-shell it is more natural to consider
the energy contribution from the localized shell. This leads to a
derivation of the Stoner-$I$ similar to the formulation of Janak
but adapted to atom-like situations.

Seo presented\cite{Seo06} the second order perturbation
theory of the spin polarization in DFT, which results
in explicit expressions for the shell exchange parameter $I_{nl}$ that are
applicable to atom like situations. In this work a numerical estimate
for $I_{nl}$ was derived indirectly from exchange splittings and
spin polarization energies taken from DFT calculations. The idea behind
this perturbation theory, the expansion the xc-energy around the
spherically averaged non-magnetic density of the shell under consideration,
was also discussed in the appendix of Kasinathan {\it et al.}\cite{Kasinathan07}
and leads to 
$\Delta E_{xc}\approx-\frac{1}{4}I_{nl}M^{2}$
with the shell-Stoner integral 
\begin{eqnarray}
I_{nl}=-\frac{1}{2\pi}\int K_{0}\left(r\right)\left[R_{nl}\left(r\right)\right]^{4}r^{2}dr
\label{eq:shellStoner}
\end{eqnarray}
\begin{eqnarray}
K_{0}\left(\vec r,\vec r'\right)&=&\left.\frac{\delta^{2}E_{xc}}{\delta m\left(\vec r\right) 
  \delta m\left(\vec r'\right)}\right|_{n^{\mathrm{spher}},m=0} \nonumber\\
 &\rightarrow& K_{0}(\vec r) \delta(\vec r - \vec r').
\end{eqnarray}
The last expression applies for a local approximation ({\it viz.} LSDA) to
$E_{xc}$. $K_{0}(\vec r,\vec r')$ is a magnetization-magnetization
interaction, directly analogous to the second functional derivative of the
DFT potential energy with respect to $n(\vec r)$, which is the Coulomb
interaction $e^2/|\vec r - \vec r'|$ plus an `xc interaction' arising from
$E_{xc}$.

For a more detailed discussion of the parameter $I_{nl}$ we performed
LSDA calculations for free atoms and ions and explicitly calculated
$I_{nl}$ from Eq. (\ref{eq:shellStoner}). It turns out that
$\Delta E_{xc}(M)$ given above is by far the largest $M$-dependent term of
the energy expansion. The spin polarization energy of isolated atoms/ions
with spherical $M$ is well described by this estimate with an error
smaller than $5-10\%$. The resulting shell-Stoner integrals $I_{nl}$
have very similar values compared to the ones obtained from the theory
for the metallic situation. (Note, however, that there is a factor
of 2 difference in the definition of the Stoner $I$ in some
of the publications.)

For the $3d$ transition element series we get values $I_{nl}$ ranging from
0.62 eV for Sc to 0.95 eV for Zn
(see Fig. 4). These values increase across the series by $\approx$ 0.15 -- 0.20 eV,
when the exchange only LSDA is used, pointing to a reduction due to
(LDA-type) correlation effects when the full xc-kernel is used. For
the $4f$ series the shell-Stoner integrals vary from 0.58 eV for Ce to
0.75 eV for Yb. The LDA correlation effects amount to $10\%$ of these
values. The values obtained depend on the choice of the reference
system, which serves as zeroth order in the functional expansion.
For instance for the $3^{+}$-ions of the $4f$-series $I_{4f}$ is
increased by $6-20\%$ with respect to the neutral atoms.

\bibliographystyle{apsrev}
\bibliography{ldau}

\end{document}